# Premelting of Al nonperfect (111) surface

F.L. Tang[*,a,b], X.G. Cheng[a], W.J. Lu[a], and W.Y. Yu[b]

[a]*State Key Lab. of Gansu Advanced Non-ferrous Metal Materials, Lanzhou Uni. of Tech.，Lanzhou 730050，China*

[b]*Key Lab. of Non-ferrous Metal Alloys and Processing of Ministry of Education，Department of Materials Science and Engineering, Lanzhou Uni. of Tech.，Lanzhou 730050，China*

**ABSTRACT**

Melting behaviors of aluminum (111) perfect/nonperfect surfaces, characterized by structure ordering parameter, have been investigated by classical molecular dynamics simulation with embedded atom method potential. Al (111) perfect surface has a superheating temperature above bulk Al melting point $T_m$, in this work, by about 80 K. Al nonperfect (111) surface has somewhat different local lattice structure from that on (111) perfect surface. Al nonperfect (111) surfaces tempt to premelt when temperature is less than $T_m$, in our simulation, by about 45 K. Aluminum atoms on the nonperfect surface zones are the sources of surface melting, and have larger velocities than those on the perfect surface zones.

[*]Corresponding author. Email address: tfl03@mails.tsinghua.edu.cn (F.L. Tang).

## 1. Introduction

Metal surfaces and their melting behaviors have attracted much attention for their rich display of interesting basic-physics problems and possible applications [1-4]. For many research fields and technological applications (catalysis, cluster deposition [5], microelectronics [6]), atomic distribution on metal (for example, aluminum) surfaces (including its nanoclusters' surface) play a fundamental role, where their surface structures and melting behaviors are of primary importance.

Metal surface melting, premelting and superheating [1] have been investigated extensively using different experimental techniques such as electron diffraction [7,8], low energy ion scattering [9,10], X-ray reflectivity [11], as well as atomic simulation [12-14]. Generally, experiments and simulations found three types of surface melting behaviors [1,15]: (1) open (110) surfaces of face-centered cubic (fcc) metals, e.g. Al [16] and Cu [17] exhibit a complete surface premelting; (2) close-packed (111) surfaces of some fcc metals (such as Al, Pb, Au, etc.) do not melt below bulk melting point $T_m$ [18] or superheat above $T_m$; (3) (100) surfaces of fcc metals such as Pb [19], Ni [20] and Au [21], exhibit incomplete surface melting, i.e. thickness of the surface liquid layer remains finite as $T_m$ is approached.

The lattice structure and high-temperature properties of Al (111) perfect surface had been studied by molecular dynamics simulation [22]. It was found that this surface does not melt below the bulk melting point, but can be superheated, evidenced in several experiments [23,24]. Like Al (111) surface, Pb (111) has a transient superheating up to 120 K above $T_m$, observed by time-resolved reflection high-energy electron diffraction [25]. Experiments found that aluminum (110) surface premelt, commencing at 150 K below Al bulk melting point $T_m$ [13]. Like Al (110) surface, Pb (100) undergoes a weak disordering introduced by surface vacancy at about 570 K, which is 30 K below Pb $T_m$ [14]. In addition, a strain-induced solid surface free energy increase/decrease takes place, favoring/disfavoring surface melting depending on the sign of strain relative to surface stress [26].

Although literatures had given plenty results from experiments, atomic simulation (especially molecular dynamics simulation) or *ab initio* calculation, focusing on the atomic structure and melting behaviors of perfect aluminum surfaces [1-6,22], there appears to be little published data for nonperfect surface melting and their local lattice structure. In this paper, we used classical molecular dynamics simulation with embedded atom method potential to study the details of surface structure and premelting of Al nonperfect (111) surfaces.

## 2. Simulation method

The crystal structure of a bulk material at a given temperature and pressure can be predicted by minimizing its free energy [27]. Our approach is to adjust the cell volume and atomic positions until the net pressure or stress is zero. Calculating the free energy at a given volume and then recalculating it after making a small adjustment to the cell volume determines the pressure. During the iterative procedure, a constant volume energy minimization is performed. Hence, each time the cell volume is modified; all atomic positions are adjusted so that they remain at a potential energy minimum. Thus the crystal structure at a given temperature and pressure can be predicted. In this work, atomistic simulation technique in the frame of embedded atom method (EAM) is used to calculate the free energy of aluminum bulk, surfaces and clusters.

In EAM, the cohesive energy of an assembly of *N* atoms is defined as

$$E_{coh} = \sum_i F_i(\rho_i) + \sum_{i>j} \phi(r_{ij}) \text{ and} \tag{1}$$

$$\rho_i = \sum_{j \neq i} f(r_{ij}), \tag{2}$$

where $E_{coh}$ is the total cohesive energy, $\rho_i$ is the host electron density at the location of atom *i* introduced by all other atoms, $f(r_{ij})$ is the electronic density of atom *i* as a function from its center, $r_{ij}$ is the separation between *i* and *j* atoms, $F_i(\rho_i)$ is the embedding energy to embed atom *i* in the electron density $\rho_i$. $\phi(r_{ij})$ is the Buckingham pairwise potential energy function between *i* and *j* Al atoms:

$$\phi(r) = A \exp(-r/\rho) - Cr^{-6}, \tag{3}$$

where $A$, $\rho$, and $C$ are fitting parameters.

We used this technique to simulate many kinds of materials [28-32]: from bulk [28,29,31] to surface [32] and to nanocluster [30]. Details of this technique are available in [27] and [31]. The potential parameters used for aluminum [33,34] can well reproduce the experimental crystal structure. The calculated aluminum lattice constant *a* is 4.0479 Å (4.05 Å [35]). The bulk module is 81.34 GPa (76.2 GPa [36]). In addition, simulated melting point of Al is 1000 K (933 K [37]). To simulate melting point, perfect lattice is used whereas there are different types of lattice defects (surface, grain boundary, vacancy etc.) in real bulk Al before melting. This is the reason why the simulated melting point is slightly larger than experimental value. These values by molecular dynamics [38,39] and EAM [40] potential are in a good agreement with the relevant experimental values (in the brackets) and give us confidence to simulate Al perfect/nonperfect surfaces.

## 3. Results and discussion

*3.1. Surface energies of aluminum perfect surfaces*

We had simulated (001), (111) and (110) surfaces of aluminum using lattice statistics method and provided their surface energies in [30]. Here the simulation procedure and results are briefly given for convenience. To obtain a suitable surface slab model, for example, (001) surface, and make the calculations most efficient, the unit cell of Al was extended to two times along the *x*, *y* axis directions and six times along the *z* axis direction. The surface slab has two-dimensional periodic boundary conditions parallel to the surface. The slab was split into two regions (Fig. 1a: I and II). Above region I, there is a semi-infinite vacuum. During the simulation, the atoms of the region I structural units were relaxed explicitly until there is zero force on each of them, whilst those in region II were kept fixed to reproduce the potential of the bulk lattice on region I. Details of this technique for surface simulation are available in [41]. The lattice parameters *a* and *b* of the slab were kept fixed during the simulation, thus the surface energy $E_s$ can be calculated [32,41] as

$$E_s = \frac{E_{slab} - mE_{bulk}}{S}, \qquad (4)$$

where $E_{slab}$ is the total energy of the two-dimensional slab with *m* Al formula units, $E_{bulk}$ is the total energy per unit of the Al bulk and *S* is the surface area of the slab.

In our simulation, six atomic layers in the surface (region I) were relaxed. It is found that the surface atoms have small relaxations compared with their bulk positions. For these three types of surfaces, all the surface layers shift inwards with displacements perpendicular to the surface. And no displacement parallel to the surface takes place. For every kind of surface, the top-layer has the largest displacement: $D_{111}$ = 1.0%, $D_{001}$ = 1.6% and $D_{110}$ = 2.3% of Al lattice constant $a$ (4.05 Å [28]) for (111), (001) and (110) surface, respectively.

Our simulated surface energies of Al (001), (111) and (110) surfaces are shown in table 1 and compared with other computational values. It is found that unrelaxed surfaces have slightly larger surface energies than those of relaxed surfaces, indicating that these surfaces undergo small surface relaxations (as shown above). Relaxed (111) surface has the smallest surface energy $E_{111}$ = 0.75 J/m$^2$, whereas relaxed (110) has the largest surface energy $E_{110}$ = 0.91 J/m$^2$. Relaxed (001) surface has the middle value ($E_{001}$ = 0.84 J/m$^2$). Our simulated surface energies are quantitatively in good agreement with other simulated values (table 1) [42-44]. Most important, our surface energy order is same as that from other simulated results.

*3.2. Surface structures of Al nonperfect (111) surfaces*

Sketch-maps of Al nonperfect surface with a pit or a plateau are shown in Fig. 1b and 1c, respectively. The nonperfect surface can be divided into two zones: the defect zone (on the plateau or near the pit) and the perfect zone. Aluminum nonperfect (111) surface model used in this paper is shown in Fig. 2a, with one plateau (two layers height) and one pit (two layers depth) on the surface. In order to simulate the Al nonperfect (111) surface by molecular dynamics, we obtained its atomic structure at 0 K temperature.

Before relaxation by lattice statistics, in another word, just after cleavage, the surface bond lengths in the same layer or the inter-layer bond lengths are 2.864 Å.

After relaxation, not only the atoms on the plateau but also the atoms near the pit attract themselves: the bond lengths on the defect zones decrease. On the plateau (Fig. 3a), the Al-Al bond length decreases from 2.864 Å (before relaxation) to the average value 2.686 Å, ranging from 2.666 Å to 2.721 Å. Near the pit (Fig. 3b), the Al-Al bond length decreases from 2.864 Å (before relaxation) to the average value 2.839 Å, ranging from 2.830 Å to 2.858 Å. Obviously, the plateau bond length has a larger contraction 6.2% after cleavage than the value of the pit bond length (0.9%). On the perfect zone, the Al-Al bond length (not shown) is affected by defect zones. The Al-Al bond lengths on the perfect zone increase 0.1% (not shown): from 2.864 Å to the average value 2.868 Å, ranging from 2.865 Å to 2.869 Å.

*3.3. Melting of Al nonperfect/perfect (111) Surface*

In order to investigate melting behaviors at Al nonperfect/perfect (111) surfaces by molecular dynamics, we constructed their models from the coordinates of three-dimension periodic Al superlattice. We then heated the surface models and the bulk Al model from 10 K to 1200 K. It is found that the resulting structures of heated Al bulk or surfaces depend on the heating rate. To reduce heating rate effect, we slowly (time step is 1 fs) increased the temperature from 10 K to 300 K in 30 ps and kept it 30 ps (equilibrium period) for equilibrium condition. Then the temperature was kept 40 ps (production period) to collect the structural data. Calculation for higher temperature was based on the production model of previous lower temperature. This can reduce heating rate effect, and can save calculation time.

During the production period, it was found that the atoms on defect zones have larger velocities than those of the atoms on perfect zones. At the end of the production period (50 steps are randomly selected and averaged), the velocity ($V_{top}$, the surface model for atomic velocity is shown in Fig. 4d) of 13 atoms on the top of the plateau (linked by the bond lengths in Fig. 3a ) and the velocity ($V_{edge}$) of 14 atoms at the edge of the pit (linked by the bond lengths in Fig. 3b) are calculated and compared with the velocity ($V_1/V_2$) of other atoms on the first/second layer. The average velocity $V_{top}$ or $V_{edge}$ along $x$, $y$, or $z$ direction (Fig. 4a-4c) is larger than the corresponding values on the perfect zone ($V_1$, $V_2$). $V_{top}$ is similar to $V_{edge}$: they have a larger difference compared with $V_1/V_2$ along $x$ or $y$ direction than that along $z$ direction. As the temperature increases (especially > 900 K), the deference between the velocities of defect zones and those of perfect zones decreases. At higher temperature (1000 K or 1050 K), few defect atoms diffuse into the surface and they have almost same velocity as that of the atoms originally on the first layer. The difference among $V_1$, $V_2$, $V_{top}$ and $V_{edge}$ at high temperature (> 950 K) indicates that the simulation time 40 ps is not enough for a totally equilibrium melting or premelting condition. However, trying to locate the global energy minimum is a far more time-consuming and challenging task and one that has no guarantee of success, except for the simplest possible cases [33,34]. In addition, we suspect that there is a thermal fluctuation in the rough surface we considered, even in the perfect (111) surface. But this is a new topic which deserves detailed investigation.

The premelting phenomenon of Al (110) surface was simulated by density-functional molecular dynamics [45] and classical molecular dynamics [46]. The former [45] provided larger MSD (mean square displacement) on (110) surface (along *x* direction, 400 K to 900 K) than that in the bulk. We suspect that the larger MSD on the surface may be contributed to a larger velocity of surface atoms than that in the bulk. The later [46] found that the MSD along *x* or *y* direction is much larger than that along *z* direction in (110) surface (250 K to 700 K). We suppose that the atomic velocity parallel to the (110) surface is larger than that perpendicular to the surface. In (111) surface, as shown in Fig. 4a-4c, atomic velocity along *z* direction is larger than that along *x* or *y* direction. This may be caused by the different distance between the first and the second layer, and different type of these surfaces: (111) is close-packed and (110) is open-packed.

To monitor the melting procedures of Al bulk and (111) surfaces, a structure ordering parameter, *SOP* is defined as [47]:

$$SOP = \frac{1}{N^2} \left\langle \sum_i \sum_j \exp(i\bar{k}\cdot r_i - i\bar{k}\cdot r_j) \right\rangle, \qquad (5)$$

where *N* is the number of atoms in the bulk or in the relaxed layers of the surface models and $\bar{k}$ is chosen to probe the nearest neighbor distance. *SOP* of Al bulk was calculated from all the atoms in the three-dimension lattice. *SOP* of nonperfect/perfect (111) surfaces was calculated from the atoms of the relaxed six layers. They are shown in Fig. 5.

Aluminum bulk exhibits a sharp phase transition from solid to liquid at 1000 K: its *SOP* (Fig. 5) decreases gradually from 0.98 to 0.55 when temperature increases from 10 K to 1000 K. Its *SOP* is almost zero when temperature larger than 1000 K. The *SOP* of Al perfect (111) surface also decreases gradually from 0.97 to 0.42 when temperature increases from 10 K to 1080 K. The temperature difference between Al perfect (111) melting point (1080 K) and Al bulk melting point (1000 K) indicates that the former has a superheating temperature about 80 K in this simulation. Bilalbegović [22] studied the structure and metastability of superheated Al (111) surface and estimated that the maximum superheating temperature is 180 K. The transition from superheated to liquid state was analyzed using the Fokker-Planck equation with some experimental data (latent heat of melting, bulk melting temperature, specific heat and diffusion constant), and the value ∼23 K was obtained for the maximum of superheating. Using the same potential from Bilalbegović [22], Di Tolla and coworkers reported a superheating temperature ∼150 K for the same surface. The difference between two superheating temperatures was contributed to different shape and size of molecular dynamics boxes.
H

From Fig. 5, it is found that the *SOP* of Al nonperfect (111) surface decreases gradually from 0.95 to 0.52 when temperature increases from 10 K to 900 K than decrease abruptly to almost zero when temperature is larger than 955 K. Compared with the melting point (1000 K) of bulk Al, this nonperfect surface has a premelting temperature 45 K in our simulation. Below premelting temperature, for example, at 750 K, the atoms on plateau (in the blue circle) or near the pit (in the green circle) in Fig. 2a have leaved away from their original positions much and melted, as shown in Fig. 2b. However, the atoms on perfect zones displaced themselves from their original positions little. So the Al nonperfect (111) surface tempts to premelt and defect zones are the source of premelting.

It is well known that (111) surface is close-packed and (110) is opened on fcc metals, and the atomic density on (100) surface is larger than that on (110) whereas smaller than that on (111). Just as indicated in Introduction, if a surface has a closer atomic arrangement, it has a higher melting point. It seems that denser the surface atoms are packed, harder the surface can be melted. We have calculated the atomic densities (the number of atoms per unit area of the surface) on Al surfaces: 0.141 atom/$Å^2$, 0.122 atom/$Å^2$ and 0.086 atom/$Å^2$ for (111), (001) and (110) crystal plane, respectively [30]. In the melting sense from our simulation, we suspect that nonpefect (111) surface has similar local lattice structure as that on (110) or (100) surfaces, at least, at the step positions of defect zones. This may be the reason why Al nonperfect (111) premelt. Hendy and Schebarchov [48,49] recently found that the premelting of (100) facets can coexist with solid (111) facets on sufficiently large aluminum nanoparticles, and the premelting of (100) facets limits superheating effect from (111) facets. Like the effect of (100) surface premelting on Al nanoparticles [48], the premelting of defect zone of Al (111) surface, prior to the melting of perfect (111) surface, ultimately introduces the premelting of the whole (111) surface.

Here we would like to give a remark on the simulation for melting of Al nonperfect (111) surface. It is obviously that only one surface model for every kind of perfect surface, and its simulated melting behaviors only depend on simulation parameters: inter-atomic potentials and simulation procedure (time step, equilibrium and production time, temperature change rate and so on). However, there are numerous types of surface configurations for every kind of surface, for example, (111) surface. Their melting behaviors depend on not only inter-atomic potentials and simulation procedure but also atomic arrangement on the surfaces. It is impossible, and not necessary, to simulate all types of nonperfect surface models. We must point out that our results just illustrate the trend of premelting behaviors on noperfect (111) surface, and that other research may obtain different or somewhat different data for different nonperfect surface models (for a simple example, different percentage of defect zones on the surface).

## 4. Conclusion

Using embedded atom method potential, we performed molecular dynamics simulation on aluminum (111) perfect and nonperfect surfaces to investigate their surface structures and melting behaviors. Specific conclusions are as follows:

(1) Different from its perfect (111) surface, Al nonperfect (111) surfaces tempt to premelt.

(2) Atoms at defect zone on Al nonperfect (111) surface have different local lattice structure and larger from those at perfect zone. Atoms at defect zone have larger velocities than those at perfect zone, and they are the sources of surface melting.

**ACKNOWLEDGMENT**

We were supported by National Science Foundation of China (10964003), Natural Science Foundation of Gansu Province (096RJZA102), and LUT Research Development Funding (BS01200905), and this work was performed in Gansu Province Supercomputer Center.

Table 1
Surface energies (J/m$^2$) of perfect (001), (111) and (110) crystal planes compared with other calculated values.

|  | Miller index | | |
| --- | --- | --- | --- |
|  | (001) | (111) | (110) |
| Unrelaxed | 0.85 | 0.76 | 0.93 |
| Relexed | 0.84 | 0.75 | 0.91 |
| [42] | 0.86 | / | 1.10 |
| [43] | 0.92 | 0.89 | 1.02 |
| [44] | 0.98 | 0.93 | / |

# Figure captions

Fig. 1. (Color online) Aluminum surface slab model (a). This surface slab model is also extended to several times along the *x*, *y* axis directions for different surface, not shown for simplicity. Sketch-maps of Al nonperfect surface with a pit (b) or a plateau (c).

Fig. 2. (Color online) Aluminum nonperect (111) surface with a pit and a plateau at 0 K after cleavage (a) and before premelting (b).

Fig. 3. (Color online) Relaxed local lattice structure of defect zones at Aluminum nonperfect (111) surface: (a) on the plateau (golden ball: Al atom on the first layer of the surface) and (b) near the pit (golden ball: Al atom on the second layer in the surface). (a)/(b) is the top view of the circle A/B in Fig. 2a after relaxation.

Fig. 4 Atomic velocities of defect zones ($V_{top}$ and $V_{edge}$) and perfect zones ($V_1$ and $V_2$) along *x* (a), *y* (b), or *z* (c) direction. (d) is the surface model for illustration of the velocities.

Fig. 5. (Color online) Structure ordering parameters of Al bulk, perfect (111) and nonperfect (111) surfaces against temperature from 10 K to 1200 K.

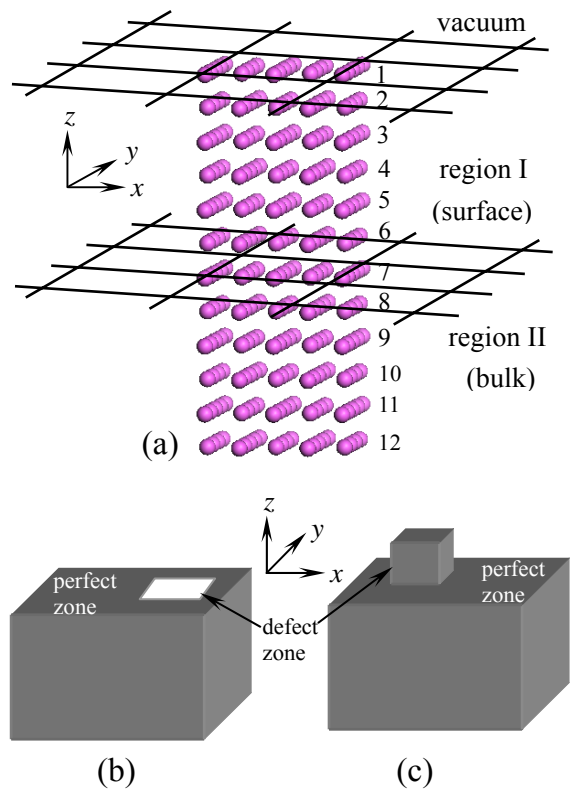

Fig. 1

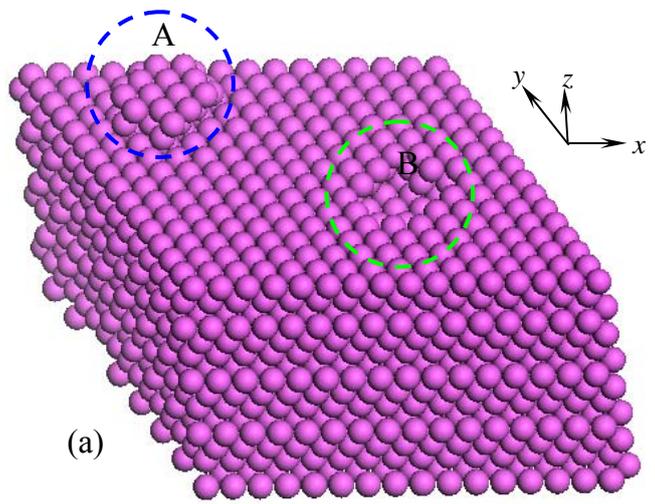

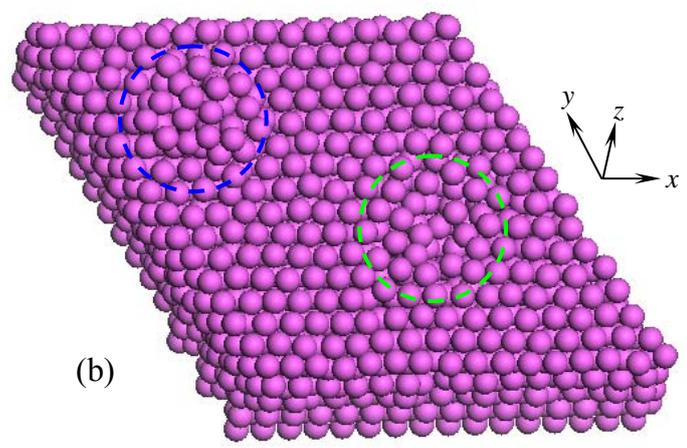

Fig. 2

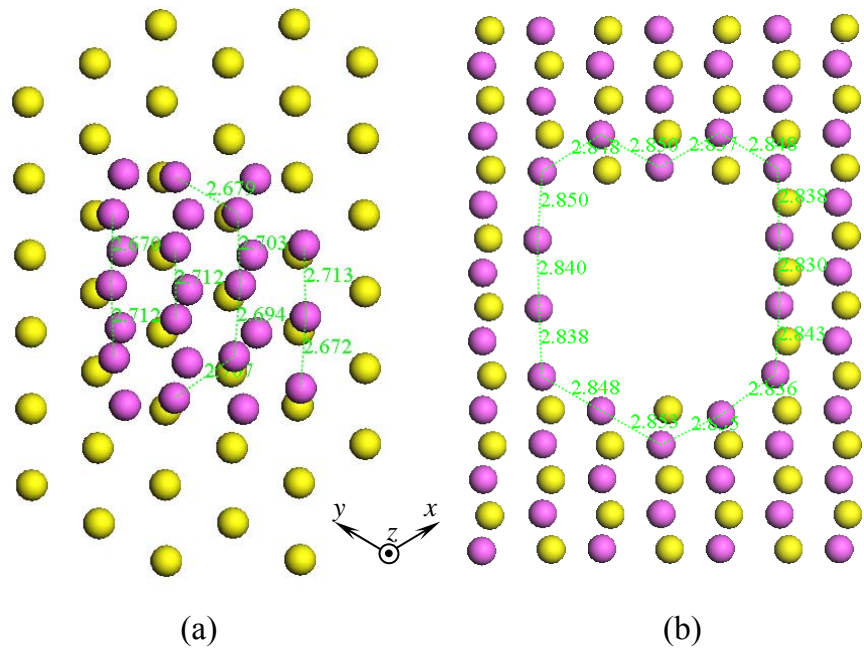

(a)                          (b)

Fig. 3

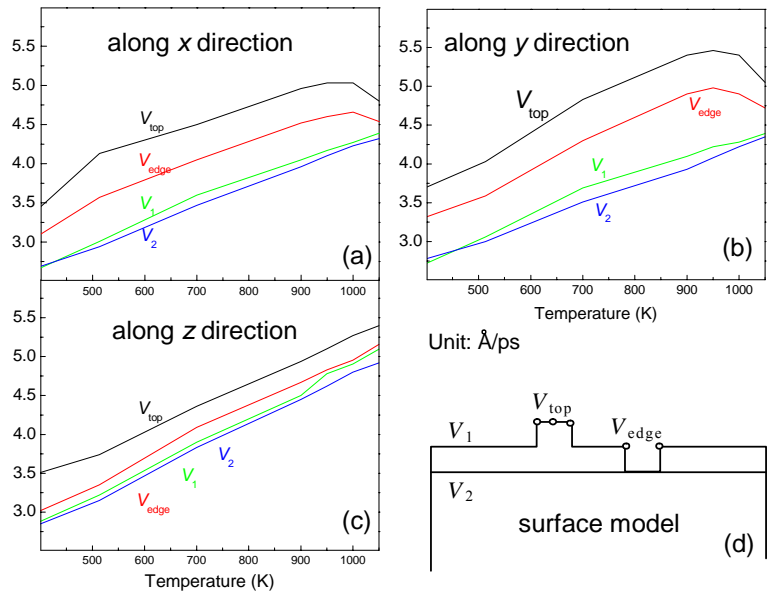

Fig. 4

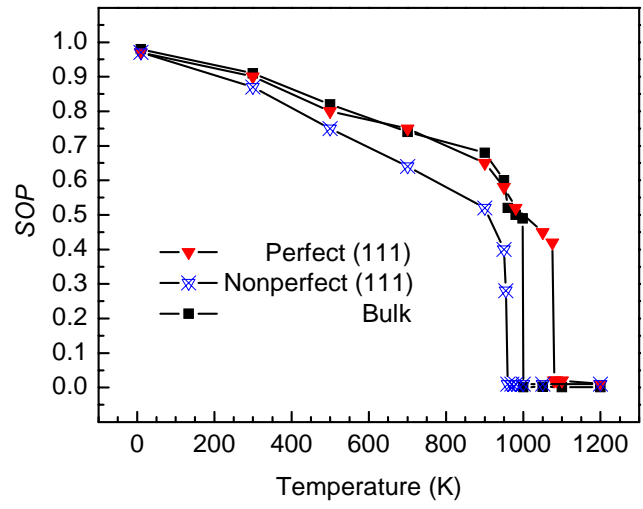

Fig. 5